\documentclass[
reprint,twocolumns,
superscriptaddress,
amsmath,amssymb,
aps,prb
]{revtex4-2}

\usepackage{amsfonts}

\usepackage{graphicx}
\usepackage{dcolumn}
\usepackage{bm}
\usepackage{hyperref}
\usepackage{physics}
\usepackage{xcolor}

\usepackage{mathrsfs}
\usepackage{mathtools}
\usepackage{comment}
\usepackage{dsfont}
\usepackage{euscript}

\newcommand{\mean}[1]{\left\langle#1\right\rangle}

\DeclareMathOperator\arctanh{arctanh}
\renewcommand{\d}{{\rm d}}

\begin{document}

\title{Semiclassical Quantum Trajectories in the Monitored Lipkin-Meshkov-Glick Model}

\author{Alessandro Santini}
\affiliation{SISSA, via Bonomea 265, 34136 Trieste, Italy}
\author{Luca Lumia}
\affiliation{SISSA, via Bonomea 265, 34136 Trieste, Italy}
\author{Mario Collura}
\affiliation{SISSA, via Bonomea 265, 34136 Trieste, Italy}
\author{Guido Giachetti}
\affiliation{LPTM, Universit\'e Paris Cergy, Av. Adolphe Chauvin 2, 95300 Pontoise, France}

\date{\today}
\begin{abstract}
    Monitored quantum system have sparked great interest in recent years due to the possibility of observing measurement-induced phase transitions (MIPTs) in the full-counting statistics of the quantum trajectories associated with different measurement outcomes. Here, we investigate the dynamics of the Lipkin-Meshkov-Glick model, composed of $N$ all-to-all interacting spins $1/2$, under a weak external monitoring. We derive a set of semiclassical stochastic equations describing the evolution of the expectation values of global spin observables, which become exact in the thermodynamic limit. Our results shows that the limit $N\to\infty$ does not commute with the long-time limit: while for any finite $N$ the esamble average over the noise is expected to converge towards a trivial steady state, in the thermodynamic limit a MIPT appears. The transition is not affected by post-selection issues, as it is already visible at the level of ensemble averages, thus paving the way for experimental observations. We derive a quantitative theoretical picture explaining the nature of the transition within our semiclassical picture, finding an excellent agreement with the numerics. 
\end{abstract}

\maketitle

\section{Introduction}


Realistic quantum systems are never perfectly isolated and interact with an external environment. The study of open systems has proven crucial for a faithful description of quantum evolution as seen from experiments. While isolated systems evolve purely unitarily according to the Schrödinger equation, the presence of an external environment induces non-unitary and intrinsically stochastic dynamics in open systems~\cite{BreuerPetruccione,Gardiner2004}. This interaction leads to phenomena such as decoherence and dissipation, providing insights into the correspondence between quantum and classical dynamics~\cite{zurek2009quantum}, and is essential for the development of quantum technologies~\cite{PhysRevA.77.012112}.

In recent years, the study of monitored systems -where a measurement apparatus acts as the environment- has gained significant interest. Quantum measurements not only provide information about the state of the system but also influence its dynamics through a stochastic back-action~\cite{Zurek1983Book,holevo2003statistical,Wiseman_Milburn_2009,Bassi2013RevModPhys}. Each realization of the measurement process defines a potential evolution of the system, referred to as quantum trajectory. The average state can be expressed as an ensemble average over these stochastic trajectories and, under generic assumptions, its evolution is governed by the Lindblad equation, which washes out much of the information about the microscopic measurement protocol~\cite{BreuerPetruccione}. 
On the other hand, quantum trajectories can be obtained by unraveling the Lindblad equation~\cite{carollo2019unraveling,Giulia2022_UnravelingDifferenti,Giulia2024_UnravelingDifferenti}, with different protocols leading to different underlying noisy dynamics such as quantum jumps or quantum state diffusion~\cite{PhysRevA.104.062212,gisin1992quantum,Ibarcq2016PRX}.
%
%
Studying the full-counting statistics of quantum trajectories gives access to physical phenomena which are generically not visible at the level of the ensemble averages ~\cite{Romito_1,Romito_2,Romito_3,delmonte2024monitored}, such as measurement-induced phase transitions (MIPT)~\cite{zabalo2020critical,shtanko2020classical,jian2020criticality,chan2019unitary,li2018quantum,li2019measurement,DeLuca2019,skinner2019measurement,fuji2020measurement,Boorman2022PRB,alberton2020trajectory,muller2021measurement,PhysRevB.108.184302,Romito2023Scipost,buchhold2021effective,coppola2022growth,10.21468/SciPostPhysCore.7.1.011,Tirrito2023Scipost,PhysRevResearch.6.023176,Gerbino2024,sharma2022measurement,turkeshi2022entanglement,PhysRevResearch.6.023176,cecile2024measurementinducedphasetransitionsmatrix,PhysRevX.13.041046,PhysRevLett.132.110403}, quantum error correction and information scrambling~\cite{PhysRevLett.125.030505,PhysRevX.10.041020}, and dynamical purification~\cite{Gopalakrishnan_2021,PhysRevB.108.L020306,giachetti2023elusivephasetransitionreplica}.

An interesting playground for the investigation of MIPTs is given by long-range interacting systems, in which the interactions between their different components decay slowly (typically as a power-law function of the distance) or do not decay at all, as in fully-connected models. Long-range physics has recently attracted a lot of attention due its peculiar properties~\cite{Defenu2023,PhysRevA.71.064101} and to the possibility of engineering such systems in experimental setups within the context of atomic, molecular and optical (AMO) systems. Examples are Rydberg atoms~\cite{LukinZollerCirac,RydbergQuantumInfo, Browaeys_2020}, trapped ions~\cite{monroe2019programmable,CiracZollerTrappedIons}, and quantum gases in a cavity~\cite{landig2016quantum,mivehvar2021cavity}, showing promising features for the development of quantum technologies. 

In this work we focus of the so-called Lipkin-Meshkov-Glick (LMG) model, i.e. a ferromagnetic all-to-all interacting spin model with $\mathbb{Z}_2$ symmetry. First introduced in the context of nuclear physics~\cite{lipkin1965validity}, the LMG model has become a paradigmatic example used to exemplify plenty of non-trivial dynamical behavior of long-range quantum systems~\cite{PhysRevB.78.104426,PhysRevE.78.021106,PhysRevB.94.184403,PhysRevB.95.214307}, such as time-crystals~\cite{PizziNatComm2021,Munozarias2022,SuracePRB2019,PhysRevB.108.L140102}, anomalous entanglement growth~\cite{pappalardi2018scrambling,lerose2018chaotic}, dynamical phase transitions~\cite{lerose2018chaotic}.  
The interplay between long-range interactions and monitoring has also been investigated ~\cite{giachetti2023elusivephasetransitionreplica,PhysRevLett.128.010604, PhysRevLett.128.010605}, finding that the range of the interaction can significantly  alter even hinder MIPTs\,\cite{minato2021fate}, as well as the effect of the global losses in AMO systems\,\cite{passerelli2024}. Moreover, the full-counting statics of the quantum trajectories in the open long-range spin systems and also in the LMG limit, has been also recently studied\,\cite{delmonte2024monitored}. 

In this paper we focus on a different class of observables. In particular, we analyse the stochastic dynamics of the expectation values of the global spin on the quantum trajectories in the thermodynamic limit, discovering that a phase transition is already present at the level of its ensemble average, a quantity which for finite size is known to converge trivially to its infinite temperature value. This signals a non-commutativity of the long-time and large-size limits, that can be traced back to the fact that, while individual quantum trajectories allow for a semiclassical description in the thermodynamic limit, this is no longer true if one consider the ensemble average of the quantum state. Let us notice that presence of a MIPT already at the presence of ensemble averages has a great interests as it allows to circumvent the post-selection issue, since these quantities are linear in the quantum state. 
%

The paper is structured as follows: after introducing the basic formalism of monitored systems in Sec.~\ref{s: monitored} and the monitored LMG model in Sec.~\ref{s: LMG_model}, we derive a closed set of coupled stochastic differential equations (SDEs) for the expectation value of the global spin over a single quantum trajectory valid in the thermodynamic limit (Sec.~\ref{s: SC lim}). We use them to quantitatively characterize the resulting MIPT in the stationary state, confronting our results with the numerics (Sec.~\ref{s:stationary}). Finally, in Sec.~\ref{s:conclusions} we draw our conclusions.

\section{Monitoring and quantum trajectories}
\label{s: monitored}
The evolution of a quantum system continuously interacting with a monitoring environment is intrinsically stochastic due to the random nature of quantum measurements. As a consequence, its time evolution is described by a stochastic Schrödinger equation (SSE). Many detection protocols exist and here we focus on the case of weak Gaussian measurements~\cite{Jacobs_2006}, which can be experimentally reproduced with a homodyne detection scheme~\cite{Wiseman_Milburn_2009, Jacobs_2006}.
Considering a pure state evolving under the action of a Hamiltonian $H$ and continuous monitoring of observable $X$ with strength $\gamma$, its SSE is given by (see App.~\ref{app:weak_monitoring} for details)
\begin{align}
\label{eq: SSE}
 \d\ket{\psi_{t}} = -iH\ket{\psi_t}\d t + \bigl(\ket{v}\gamma \d t + \ket{u} \sqrt{\gamma} \,  \d \xi\bigr),
\end{align}
where $\d \xi$ is a real It\=o differential with $\overline {\d \xi^2} = \d t$ and
\begin{align}
    \ket{v} = -\frac{1}{2}(X-\mean{X})^2\ket{\psi_t},&&
    \ket{u} = (X-\mean{X})\ket{\psi_t}.
\end{align}
The SSE is solved by an ensemble of quantum trajectories $\{|\psi(t)_\xi\rangle\}$ in the Hilbert space, each induced by a specific realization of the white noise $\d \xi$. Correspondingly, the expectation values of an observable $O$ evaluated along a trajectory evolve along a classical stochastic process. According to It\=o
calculus~\cite{gardiner2009stochastic}, we expand
\begin{equation}
    \d\mean{O} = \mel{\d\psi}{O}{\psi} + \mel{\psi}{O}{\d\psi} + \mel{\d\psi}{O}{\d\psi}\,,
\end{equation}
where $\mean{O} = \mel{\psi}{O}{\psi}$ and $ \mel{\d\psi}{O}{\d\psi} =\gamma \d t \expval{O}{u}$ as a consequence of $\d \xi^2 \approx \d t$. The stochastic differential equation satisfied by $\mean{O}$ is then
\begin{equation}
\label{eq: SHE}
\begin{split}
     \d\mean{O} = i\d t \mean{\comm{H}{O}}  &+ \mean{\acomm{X-\mean{X}}{O}} \sqrt{\gamma} \d \xi \\
   - \frac{\gamma}{2}  \d t & \mean{\comm{X}{\comm{X}{O}}}.
\end{split}
\end{equation}

\begin{figure}[t]
    \centering
    \includegraphics[width=\linewidth]{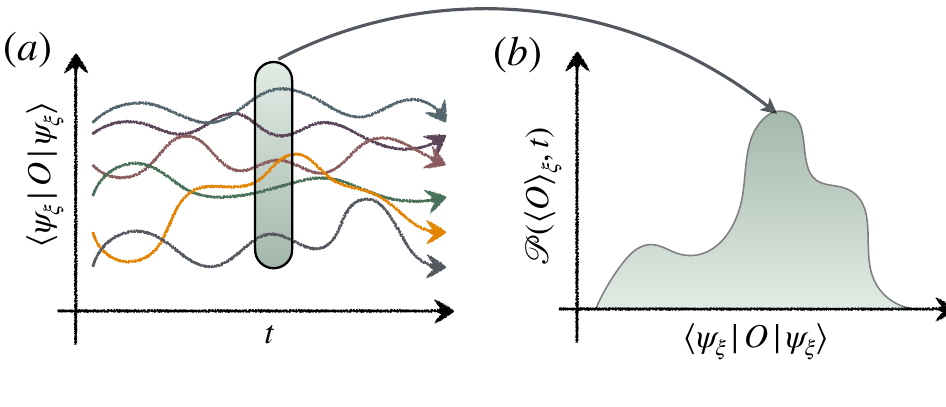}
    \caption{(a) Evolution of the expectation value of a set of quantum trajectories. (b) Probability distribution of the expectation values of the observable $ O$ over the ensemble of quantum trajectories. }
    \label{fig:scheme_lindblad}
\end{figure}

\subsection{Ensemble averages and full counting statistics}
\label{s:unraveling}
Let us introduce the average of the state of the system over all measurement outcomes, namely $\rho = \overline{\dyad{\psi_\xi}}$, where $\overline{\cdots}$ denotes the average over $\d \xi$. From Eq.~\eqref{eq: SSE} it is immediate to find that the evolution over time of $\rho$ is described by 
\begin{equation}
\label{eq: lindblad}
    \dv{\rho}{t} = -i \comm{H}{\rho} - \frac{\gamma}{2} \comm{X}{\comm{X}{\rho}}\,,
\end{equation}
that is a master equation of the Lindblad form. The observable $X$ is playing the role of the jump operator, which encode the interaction with the external environment. As the jump operator is Hermitian, for any finite dimension $N$ the Linblad dynamics will evolve towards the maximally mixed state in any of the symmetry sector of the Hilbert space, thus washing away all the information contained in the initial state $\rho(0)$ except for the presence of invariant subspaces~\cite{Wiseman_Milburn_2009,Nielsen}. If no symmetry in present then $\rho(t\to\infty) \propto \mathbb{I}$ over the whole Hilbert space.  

In the same way, the averages along trajectories of expectation values follow Lindblad dynamics since $\overline{\mean{O}} = \overline{\text{tr}[\dyad{\psi_\xi}O]} = \text{tr}[\rho \, O]\equiv\mean{O}_\rho$. Their evolution is generated by the adjoint Lindbladian 
\begin{equation}
\label{eq: adjoint lindblad}
     \d\overline{\mean{O}} = i\d t \overline{\mean{\comm{H}{O}}} - \frac{\gamma}{2}  \d t \overline{\mean{\comm{X}{\comm{X}{O}}}}\,.
\end{equation}
Analogously, for finite $N$ we have that $\overline{\mean{O}}$ will converge asymptotically to its infinite-temperature expectation values for large times. 

While linear averages only admit trivial long-time asymptotic values, it is important to note that important features of the ensemble of quantum trajectories are hidden in the higher moments of the distribution. A prominent example is the purity: starting from a pure state, while for any $t > 0$ $\rho(t)$ will be a mixed state, asymptotically reaching the maximally mixed state, individual trajectories remain pure, so that 
\begin{equation}
    \Tr{\rho^2} \leq \overline{\Tr{\dyad{\psi_\xi}^2}}=1\,.
\end{equation}
In other words, the SSE carries more information than the Lindblad equation. 
Here, we want to go beyond the Lindblad approach and study statistical properties of the ensemble of weak monitoring quantum trajectories. As an example, as depicted in Fig.~\ref{fig:scheme_lindblad}, we consider the characteristics of the probability distribution of the expectation values of operators~\cite{Tirrito2023Scipost,Lami_2024,Giulia2023PRB_LongRange,turkeshi2024density} \begin{equation}
    P(x,t) = \overline{\delta(x-\expval{O}{\psi_\xi})}\,.
\end{equation}

\section{Monitored LMG Model}
\label{s: LMG_model}

Given a set of $N$ spin-$1/2$ particles, each located in a lattice site $j$, the LMG model is described by the following Hamiltonian
\begin{equation}
\label{eq: H_LMG}
    H = -\frac{1}{2N}\sum_{ij} \sigma^x_i\sigma^x_j -h\sum_i \sigma^z_i,
\end{equation}
where $\sigma_j^\alpha$ are the Pauli operators for each site, $h$ is a magnetic field and the $1/N$ factor is the Kac scaling~\cite{kac1963van} that ensures the correct extensive scaling of the energy. We are working in units such that $\hbar =1$. The above Hamiltonian can be seen as a quantum Ising model in transverse field, in presence of infinite-range ferromagnetic interactions. It is convenient to express it in terms of collective spin operators $S_\alpha$ or, equivalently, of the reduced global magnetization $\hat{m}_\alpha$
\begin{equation}
    S_\alpha = \frac{1}{2}\sum_j \sigma^\alpha_j\,\,,\,\,\,\,
    \hat{m}_\alpha = \frac{S_\alpha}{S} = \frac{1}{N}\sum_j \sigma^\alpha_j \,,
\end{equation} 
with $\alpha =x,y,z$, $S=N/2$, so that 
\begin{equation} \label{eq: HLMG m S}
    H = -\frac{1}{S}S_x^2-2hS_z = -\frac{N}{2}\left(\hat{m}_x^2 + 2h\hat{m}_z \right)\,. 
\end{equation}
As $H$ and $\mathbf{S}^2$ commute, the unitary dynamics takes place in subspaces with fixed $\mathbf{S}^2 = S(S+1)$, $S=1, \dots, N/2$.  

Let us now introduce a deformation of the the LMG model in which the global spin $S_z$ undergoes a continuous weak monitoring. Due to the global nature of the montiored obsevables, the dynamics of the resulting SSE~\eqref{eq: SSE} will share the same symmetry of the unitary evolution, so that the monitored dynamics will also decompose into invariant subspaces with $\mathbf{S}^2 = S(S+1)$ fixed. In particular we choose the initial state to be fully-magnetized along a specific spatial direction (not necessarily coinciding with the $z$-axis), so that the dynamics will be confined on the subspace $S=N/2$ ($\mathbf{\hat{m}}^2 = (S+1)/S$). Such sector corresponds to the representation of spin $N/2$ of the total angular momentum, which is totally symmetric with respect to the permutation symmetry of the LMG model.
We will focus on the statistics of the expectation values of the global magnetization $\hat{m}_\alpha$: in this case Eq.~\eqref{eq: SHE} takes the form
\begin{equation}
\label{eq: SHELMG}
\begin{split}
    d \mean{\hat{m}_x} &= 2 h \mean{\hat{m}_y} \d t - \frac{\gamma}{2} \mean{\hat{m}_x} \d t + \sqrt{\gamma} S \d \xi \mean{\hat{m}_z,\hat{m}_x}_c \, ,  \\
    d \mean{\hat{m}_y} &= \mean{ \acomm{\hat{m}_x}{\hat{m}_z} - 2h \hat{m}_x} \d t - \frac{\gamma}{2} \mean{\hat{m}_y}  \d t \\ 
    &+ \sqrt{\gamma} S \d \xi \mean{\hat{m}_z,\hat{m}_y}_c \, , \\
    d \mean{\hat{m}_z} &= - \mean{ \acomm{\hat{m}_x}{\hat{m}_y}} \d t - \frac{\gamma}{2} \mean{\hat{m}_z} \d t \\ 
    &+ \sqrt{\gamma} S \d \xi \mean{\hat{m}_z,\hat{m}_z}_c \, , \\ 
\end{split}
\end{equation}
where $\mean{A,B}_c \equiv \mean{AB+BA} -  2 \mean{A} \mean{B}$.

\section{Semiclassical limit}
\label{s: SC lim}

The unitary LMG model in the thermodynamical limit behaves semiclassically. In particular, collective spin operators in the large-$N$ limit can be seen  as classical angular momentum variables. Formally, this is due to the fact that the ${\rm SU}(2)$ algebra implies
\begin{equation}
\label{eq: commutator}
[\hat{m}_\alpha,\,\hat{m}_\beta] = \frac{i}{S} \epsilon_{\alpha\beta\gamma} {\hat{m}_\gamma}\,, 
\end{equation}
so that $1/S$ plays the role of an effective Planck constant. Therefore, for $S\gg1$ there will be states for which the uncertainty over all the components of $\hat{m}_\alpha$ is simultaneously small (i.e. $O(1/S)$). 
On such states one has
\begin{equation}
\label{eq: large S approx}
    \langle \hat{m}_\alpha \hat{m}_\beta \rangle
= \langle \hat{m}_\alpha\rangle \langle \hat{m}_\beta \rangle + O(1/S)\,,
\end{equation}
which is key for the semiclassical description. 
This is the case for the dynamics in the totally symmetric sector, where $\mean{S_\alpha} = O(N)$ and the uncertainty is of order $O(1)$. In this sector, we can approximate the state with a coherent spin state of spin $S=N/2$, i.e. the eigenstate of the projection of the spin operators along the direction of $\langle \mathbf{S}\rangle = (\mean{S_x},\mean{S_y},\mean{S_z})$. 
A summary of the properties of coherent spin states is given in App.~\ref{app:coherentstates}. 
Let us notice that while the mean field approximation works along individual trajectories, as we shall see in Sec.~\ref{s: monitored semiclassical}, it is not reliable for the Lindblad dynamics of the mean state. Taking ensemble averages, 
\begin{equation*}
\langle \hat{m}_\alpha \hat{m}_\beta \rangle_\rho = \overline{\langle \hat{m}_\alpha \hat{m}_\beta \rangle} \neq \overline{\langle\hat{m}_\alpha\rangle}\,\overline{\langle\hat{m}_\beta\rangle} = \langle \hat{m}_\alpha\rangle_\rho \langle\hat{m}_\beta\rangle_\rho\,.
\end{equation*}
Indeed, while the expectation value $\mean{\cdot} = \tr{  \dyad{\psi_\xi} \cdot }$ is performed on a pure state with $\mean{S_\alpha} = O(N)$, which can be well approximated with a spin coherent state, for the ensemble $\overline{\mean{\cdot}} = \langle \cdot \rangle_\rho = \tr{ \rho \ \cdot }$ the quantum expectation value is computed on the highly mixed state $\rho$, that has no well-defined mean-field limit. This is apparent for long times, as the Lindblad dynamics will bring $\rho(t)$ towardss the maximally mixed state in the totally symmetric subspace, so that $\overline{\langle \hat{m}_\alpha\rangle} = 0$, while  $ \overline{\langle \hat{m}_\alpha \hat{m}_\beta \rangle} = (S+1)/S \cdot \delta_{\alpha, \beta}$. For more details we refer the reader to App.~\ref{app: semiclassical approximation}.

We will now briefly review the consequences of the mean field approximation on the unitary dynamics, before extending the description to the monitored case.
 
\begin{figure*}[t]
    \centering
\includegraphics[width=.9\linewidth]{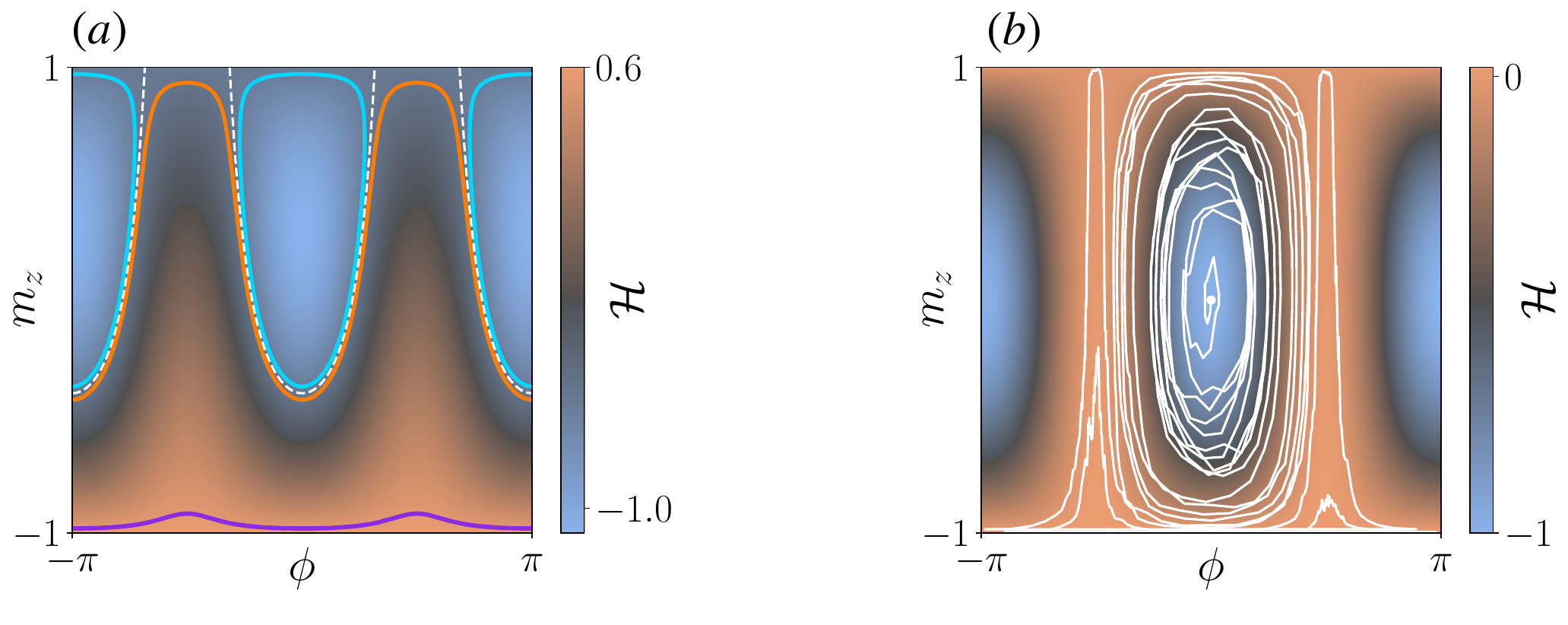}
    \caption{ (a) Energy landscape of the unitary dynamics ($\gamma = 0$) for $h=0.3$. As $h<1$, a separatrix $\mathcal{H}(m_z, \phi) = - 2h$ is present (dashed white line). Trajectories with energy $ < - 2h$ correspond to librations (light blue curve), otherwise to rotations (orange and purple curves). While in the vicinity of $m_z=-1$ the Hamiltonian flow remains close to the barrier during the whole dynamics (purple curves), it will eventually drive the system away from $m_z=1$, as the light blue and orange curves are close to the separatrix.  (b) Possible realization of a quantum trajectory in the phase space for small $\gamma = 0.01$  (white solid line), for $h=0.02$ and initial conditions $m_z(0)=\phi(0)=0$ ($m_y(0)=0$, $m_x(0)=1$) superimposed to the Hamiltonian energy landscape of the system, density plot. Parameters  $\gamma = 0.01$. As the noise is relatively small, on short timescales the evolution of the system remains close to the Hamiltonian flow. In spite of this, as $\gamma < \gamma_c(h) = 0.28$, the $m_z=1$ barrier is repulsive, and the trajectory converges asymptotically to $m_z=-1$.} 
\label{fig:Hamflow}
\end{figure*}

\subsection{Unitary Evolution}
Starting from a maximally magnetized state, the unitary evolution of the magnetization is given by the Eqs.~\eqref{eq: SHELMG} with $\gamma = 0$, that become a coupled system of ODEs for $m_\alpha\equiv\langle\hat{m}_\alpha\rangle$
\begin{equation}
\label{eq:classical_evo_mag}
    \dot{\mathbf{m}} = -2(m_x\hat{\mathbf{x}}+h\hat{\mathbf{z}})\times \mathbf{m}
\end{equation}
thanks to the decomposition of Eq.~\eqref{eq: large S approx}. The approximation is consistent if in the large $S$ limit the state remains coherent, which is true because $|\mathbf{m}|^2$ is conserved and the state is fully polarized. To simplify the problem we can introduce cylindrical coordinates 
\begin{equation}\label{eq:ChangeOfVariables}
	m_x= \sqrt{1-m_z^2}\cos\phi \hspace{0.15cm},\hspace{0.3cm} m_y = \sqrt{1-m_z^2}\sin\phi
\end{equation}
with $m_z\in[-1,\ 1]$ and $\phi \in [-\pi,\ \pi]$, which are convenient because $\phi$ is canonically conjugate to the third component of the angular momentum $m_z$. 
This can be checked by noticing that the equations of motion Eqs.~\ref{eq:classical_evo_mag} become Hamilton equations $\dot{m_z} = - \partial_\phi \mathcal{H}$,  $\dot{\phi} = \partial_{m_z} \mathcal{H}$, where 
\begin{equation} \label{eq: classicalHLMG}
	\mathcal{H} (m_z,\phi) = -(1 - m_z^2) \cos^2 \phi - 2 h m_z
\end{equation}
is the classical Hamiltonian (which corresponds to the classical limit of Eq.~\eqref{eq: HLMG m S} expressed in terms the new variables). Quenches of the LMG model undergo a dynamical phase transition. The dynamics will take place on the level curves $\mathcal{H}(m_z,\phi)=E$ and, as those curves are closed, the dynamics will in general be periodic. However, similarly to the classical pendulum case, we can have two different kind of dynamics. While for $h>1$ $\phi$ increases by $2 \pi$ over a period regardless of the initial condition (rotation); if $h < 1$ for $E < -2h$, $\phi$ is bounded to oscillate between two extremal values, $|\phi| < \phi_{\rm max} \ (\text{mod} \ \pi)$ (libration). In latter case the time average of $m_x$ on a trajectory is $\neq 0$ (dynamical ferromagnet). The two classes of trajectories are divided by the separatrix $\mathcal{H}(m_z,\phi) = - 2h$, i.e. 
\begin{equation} \label{eq:LMGseparatrix}
    (1+m_z) \cos^2 \phi = 2 h \hspace{1cm} |\cos \phi| > \sqrt{h} \ ,
\end{equation}
in correspondence of which the period of oscillation diverges. As a consequence, while for $h > 1$ the Hamiltonian flow remains close to the absorbing barriers $m_z= \pm 1$ for an infinite amount of time, for $h < 1$ the dynamics around $m_z=1$ takes place in proximity of a separatrix, so that it runs away from the absorbing barrier on a logarithmic timescale (see Fig.~\ref{fig:Hamflow}(a)).

\subsection{Monitored Evolution}
\label{s: monitored semiclassical}
\begin{figure*}[t]
    \centering
\includegraphics[width=0.85\linewidth]{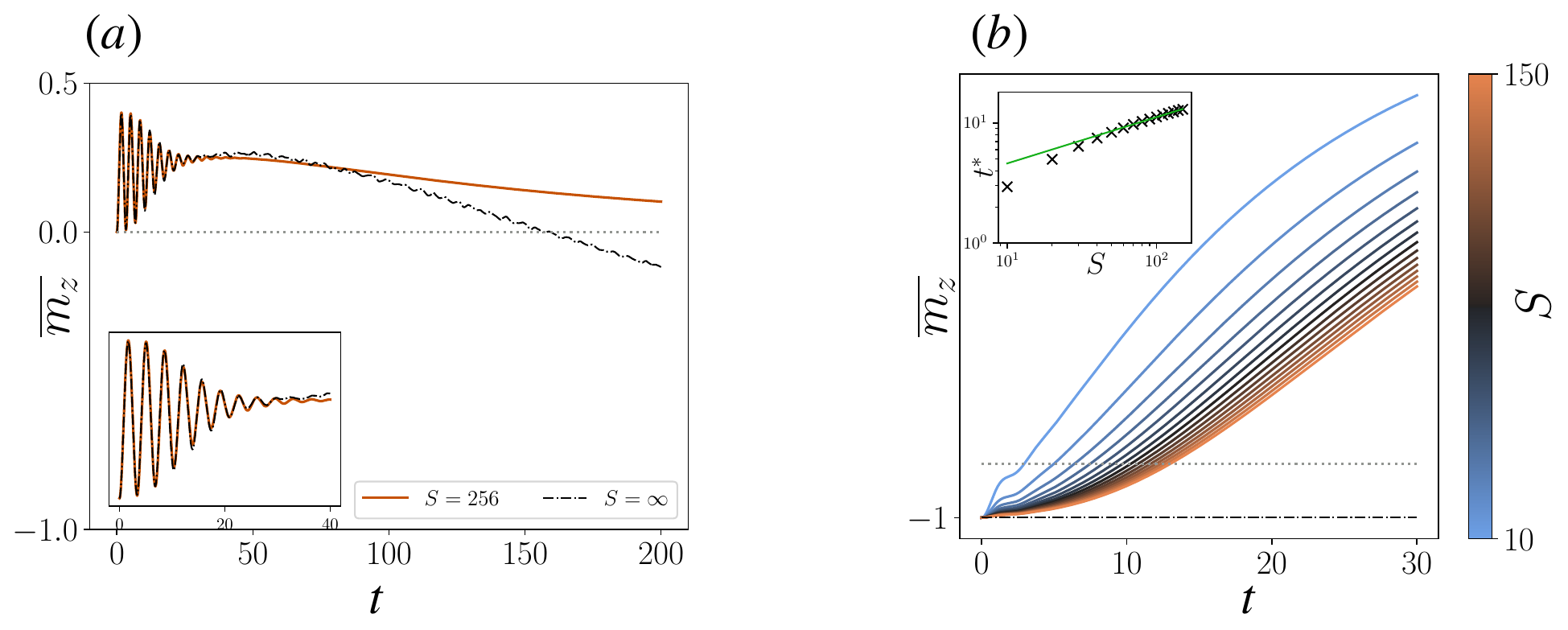}
\caption{Comparison between $\overline{m_z} (t)$ for $N= \infty$, obtained averaging over the realizations of Eq.\,\eqref{eq:LMGhamilton+noise} (black dash-dotted line), and its finite $N$ counterpart, obtained by integrating the Lindbald dynamics.
While for finite size $\overline{m_z} (t)$ goes to zero at large times, this is no longer true for $N = \infty$.
(a) $S = \infty$ and $S=N/2 = 256$ dynamics for $h=0.2$, $\gamma = 0.01$ and initial conditions $m_z(0)=m_y(0)=0$, $m_x(0)=1$. (b) Dynamics starting from the $S = \infty$ absorbing state $m_x(0)=m_y(0)=0$, $m_z(0)=-1$, compared with finite size dynamics $S = N/2 \in [10,150]$, for $h=0.3$, $\gamma=0.25$. As shown in the inset, the timescale $t^{*}$ on which the finite size dynamics differs significantly from the $S = \infty$ one (formally $m_z(t^{*}) = -0.9$) grows polynomially in $S$.  
    } 
\label{fig: largeNlarget}
\end{figure*}
In analogous fashion, we can describe the monitored LMG model in the thermodynamic limit by applying the semiclassical approximation to Eqs.~\eqref{eq: SHELMG}. Let us notice, however, that the effect of the limit $S\to\infty$ here is more subtle due to the presence of the connected correlators $\mean{\hat{m}_\alpha, \hat{m}_\beta}_c$ which vanish at the leading order if we factorize the expectation values. We thus need the exact form of the first order correction in $O(1/S)$ of Eq.~\eqref{eq: large S approx} on a coherent spin state, which is given by 
\begin{equation} \label{eq: mean_field_correction}
    \mean{\hat{m}_\alpha \hat{m}_\beta}_c = \frac{1}{S} \left( \delta_{\alpha, \beta} -\mean{\hat{m}_\alpha} \mean{\hat{m}_\beta}\right) + O(S^{-2}) \,
\end{equation}
(see App.~\ref{app:coherentstates}). By introducing once again the shorthand $m_{\alpha} = \mean{m_{\alpha}}$, for $S \gg 1$ we get another closed system of SDEs, analogous to Eq.\,\eqref{eq: SHELMG}
\begin{subequations}
\label{eq: SHE large S}
    \begin{align}
        &\d m_x = \left(  2 h m_y - \frac{\gamma}{2} m_x\right) \d t  - \sqrt{\gamma} \d \xi  m_z m_x\\
        &\d m_y = \left( - 2 h m_x + 2 m_x m_z  - \frac{\gamma}{2} m_y \right) \d t\ + \notag\\  &\qquad\qquad\qquad\qquad\qquad\qquad - \sqrt{\gamma} \d \xi m_z m_y\\
        &\d m_z =  - 2 m_x m_y \d t  + \sqrt{\gamma} \d \xi (1-m_z^2)\,.
    \end{align}
\end{subequations}
The equations imply
\begin{equation}\label{eq:evo_modulus_m_monitored}
  \d\mathbf{m}^2=  \left[\gamma  \d t \left(m_z^2-1\right)-2 m_z \sqrt{\gamma  }\d \xi \right] \left(\mathbf{m}^2-1\right)\,,
\end{equation}
so that if the initial state is chosen in the totally symmetric sector ($|\mathbf{m}|=1$) the dynamics takes place on the unit sphere and the semiclassical approximation is consistent. This is expected, as in the large $S$ limit, on the totally symmetric subspace $\mathbf{m}^2 \sim \mean{\hat{\mathbf{m}}^2} = (S+1)/S \sim 1$. 

As the dynamics takes place on the surface of the sphere, it is convenient to rewrite Eqs.\,\eqref{eq: SHE large S} in terms of $(m_z,\phi)$ defined in Eq.~\eqref{eq:ChangeOfVariables}, thus finding 
\begin{equation} \label{eq:LMGhamilton+noise}
    \begin{split}
        \d m_z &=  - 2 (1-m_z^2) \sin \phi \cos \phi \ \d t + \sqrt{\gamma} (1-m_z^2) \d \xi \\
        \d \phi &=  2(- h +  m_z \cos^2 \phi) \d t \ .
    \end{split}
\end{equation}
Notice that $m_z = \pm 1$ act now as absorbing barriers, since there $\d m_z = 0$  on each realization. As expected, for $\gamma = 0$ the evolution reduces to the Hamiltonian dynamics of the unitary case, namely
$\d m_z =  - \partial_{\phi} \mathcal{H} \d t + \sqrt{\gamma} (1-m_z^2) \d \xi$, $\d \phi = \partial_{m_z} \mathcal{H}  \d t$,
with $\mathcal{H}(m_z, \phi)$ of Eq.~\eqref{eq: classicalHLMG}. A possible realization of the noisy dynamics in the phase space with small $\gamma$ is shown in Fig.~\ref{fig:Hamflow}(b).


\subsection{Non-commutativity of the limits}
\label{s: noncommutativity}

As already mentioned, for any finite $N$ the Lindblad evolution of a maximally magnetized initial condition will converge towards the infinite-temperature state of the maximal representation $S = N/2$. This implies 
\begin{equation}
\begin{split}
 \lim_{N \rightarrow \infty} \lim_{t \rightarrow \infty} \overline{m_{\alpha}} &= 0 \,.
\end{split}
\end{equation}
On the other hand, the evolution of $m_\alpha$ in the thermodynamic limit is given by Eq.~\eqref{eq: SHE large S} which, as already noticed, has two absorbing states in correspondence of $m_z = \pm 1$ (and thus $m_x = m_y = 0$). This means that at large times each realization of the system will end up in one of the two. Depending on the initial state one of the two boundaries will be reached more easily, implying that 
\begin{equation}
     \lim_{t \rightarrow \infty} \lim_{N \rightarrow \infty} \overline{m_z} \neq 0 \, , 
\end{equation}
so that the large $N$ limit will not in general commute with the large time limit. 

For finite $N$, the Eqs.\,\eqref{eq:LMGhamilton+noise} are valid up to a timescale $t_{\rm Ehr}$, known as Ehrenfest time, after which the approximation \eqref{eq: large S approx} is known to break down\,\cite{pappalardi2018scrambling} (see also Fig.\,\ref{fig: largeNlarget}(a)). As $t_{\rm Ehr}$ can be seen as the timescale on which a phase space packet of initial width $1/S$ acquires a $O(1)$ spreading, its scaling of $t_{\rm Ehr}$ with $N$ depends strongly on the nature underlying the semiclassical dynamics. While in the chaotic case $t_{\rm Ehr} \sim \ln N$, Eqs.\,\eqref{eq:LMGhamilton+noise} describe an integrable dynamics perturbed by the presence of noise, so that $t_{\rm Ehr}$ is expected to grow polynomially $N$. This is confirmed by the numerical analysis reported in Fig.\,\ref{fig: largeNlarget}(b), where the interplay between the large-$N$ limit and the long-time limit is investigated. 


\section{Stationary probability distribution} \label{s:stationary}

\begin{figure}[t]
    \centering
    \includegraphics[width=\linewidth]{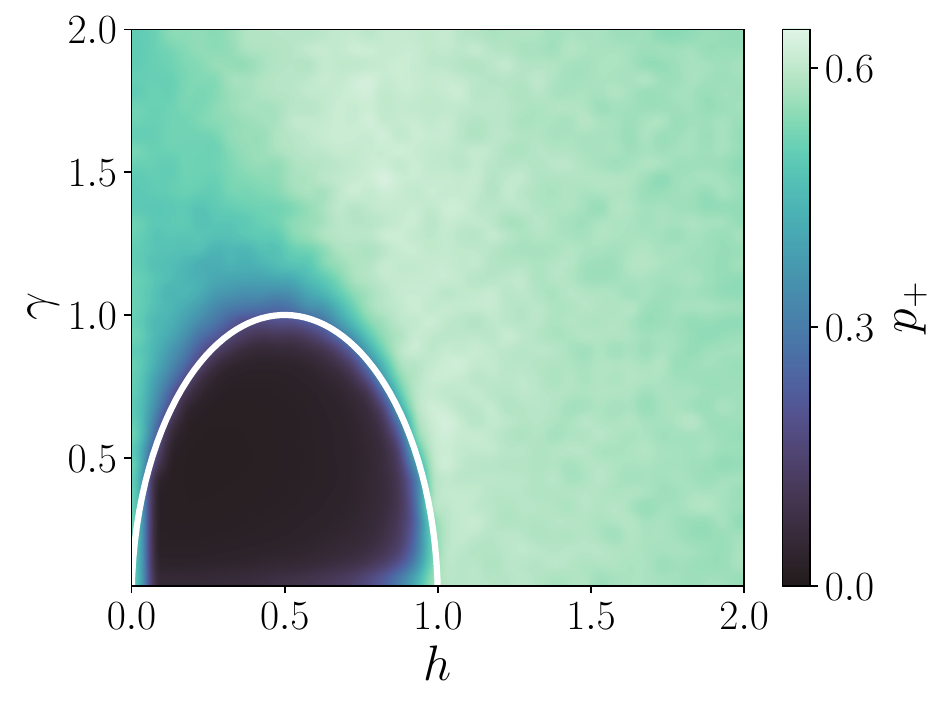}
    \caption{Color plot of $p_{+}$ as defined in Eq.~\eqref{eq:p+} with initial conditions $m_z(0)=\phi(0)=0$ ($m_y(0)=0$, $m_x(0)=1$). The phase $p_+=0$ in which only one absorbing barrier is attractive is clearly visible (black region), and it matches the theoretical estimate $\gamma < \gamma_c (h) \equiv 2 \sqrt{h(1-h)}$ (white line). While the form of such region is independent on the initial conditions, the value of $p_+$ for $\gamma > \gamma_c (h)$ will in general depend on them. The possibility of observing a phase transition by looking at observable averages is a direct consequence of the non-commutativity of the limits examined in Sec.~\ref{s: noncommutativity}.}
    \label{fig:variance}
\end{figure}

In the next section, we will analyze more closely the asymptotic distribution of $m_z$. Because of the presence of two absorbing walls, the probability distribution of $m_z$ will asymptotically have the form 
\begin{equation}
    P_{\infty} (m_z) = p_{+} \delta(m_z-1) +  p_{-} \delta(m_z+1)  \, ,
\end{equation}
with $p_{+} + p_{-} = 1$. In this case, two different scenarios are possible: either both walls can be reached with finite probability, or one of the two is forbidden by the dynamics. A good order parameter capable of distinguishing the two phases apart is thus $p_{+}$, which can be expressed in terms of the ensemble averages of $m_z$ as
\begin{equation} \label{eq:p+}
p_{+} =  \lim_{t \rightarrow \infty} \lim_{N \rightarrow \infty}  \frac{1}{2}(1 + \overline{m_z}(t)) \, . 
\end{equation}
The resulting phase diagram as function of $h$ and $\gamma$ is plotted in Fig.~\ref{fig:variance}: in the $h<1$ region, a finite phase with $p_+ = 0$ is indeed present for small $\gamma$. The boundary of the two phases meets $\gamma =0$ at $h=0$ and $h=1$.
Notice that along the Linblad evolution one would identically have $p_{+} = 1/2$, as $p_+$ is linear in the state so that the possibility of using such a quantity as an order parameter is a direct consequence of the non-commutativity of the limits explored in Sec.~\ref{s: noncommutativity}.

The expected phase diagram is confirmed by an analysis of Eqs.~\eqref{eq:LMGhamilton+noise}. 
The effect of the noise becomes crucial in the proximity of $m_z= \pm 1$, as it can push the dynamics towards the absorbing barriers. Remarkably, as in this regime both the Hamiltonian and the noisy dynamics slow down, we can treat their effects independently. In particular, close to the $m_z=-1$ barrier, the Hamiltonian dynamics is always a rotation and it remains close to the barrier at any time, so that we can ignore it. Let us denote the distance between $m_z$ and the barrier as $\Delta z \equiv 1 + m_z$. For $\Delta z \ll 1$ one has
\begin{equation}
    \d(\Delta z) = 2 \sqrt{\gamma} \Delta z \ \d \xi
\end{equation}
or, by introducing $\Delta z \equiv e^{-s}$
\begin{equation}
    \d s = 2 \gamma \d t + 2 \sqrt{\gamma} \ \d \xi \, .
\end{equation}
It follows that $s$ is a Gaussian variable centered around $2 \gamma t$ with a variance $ \sigma^2 = 4 \gamma t $: in this case the presence of the noise effectively pushes the dynamics towards the absorbing barriers at all measurement strengths. 

Close to the barrier in $m_z=1$ an analogous line of reasoning can be followed for $h>1$. For $h<1$, however, the Hamiltonian dynamics cannot be neglected since in this case the flow runs close to the separatrix \eqref{eq:LMGseparatrix} (see Fig.~\ref{fig:Hamflow}), which pushes the system away from the barrier. We thus expand the equation for $\d m_z$ in Eqs.~\eqref{eq:LMGhamilton+noise} for small $\Delta z \equiv 1 - m_z\ll 1$, assuming that $\phi$ and $m_z$ close to the separatrix that links them as in Eq.~\eqref{eq:LMGseparatrix}. At order $O(\Delta z)$ therefore we find
\begin{equation}
    \d(\Delta z) = 4 \sqrt{h(1-h)} \Delta z \d t + 2 \sqrt{\gamma} \Delta z \ \d \xi \, ,  
\end{equation}
or. in terms of $\Delta z \equiv e^{-s}$, 
\begin{equation}
    \d s = 2 \left( \gamma - 2 \sqrt{h(1-h)} \right) \d t + 2 \sqrt{\gamma} \ \d \xi  \, . 
\end{equation}
It follows that the $m_z=1$ barrier is attractive for the dynamics only when $\gamma > \gamma_c (h) \equiv 2 \sqrt{h(1-h)}$. An alternative argument is given in App.~\ref{app:separatrix}. For $\gamma = \gamma_c(h)$ the evolution of $\d s$ becomes a simple Wiener process. This means that a trajectory that starts close to the barrier $m_z=1$ has a probability $1/2$ to remain close to it for large times. We thus expect $p_+$ to be finite for $\gamma = \gamma_c(h)$, signaling a discontinuous phase transition. 

In summary, for $\gamma > \gamma_c$ both barriers are attractive, so that $p_+,\,p_-$ will both be finite, while the $m_z=1$ barrier becomes repulsive for $\gamma < \gamma_c (h)$, so that $p_{+}=0$ (see also Fig.~\ref{fig:Hamflow}(b)). This explains the numerical phase diagram of Fig.~\ref{fig:variance} even at the quantitative level, as the transition curve $\gamma_c (h)$ we foresee is in very good agreement with the numerical results. Notice also that $\gamma_c (h)$ does not depend on the specific initial condition chosen for our system, while the specific values of $p_{+}$ within the $p_{+} \neq 0$ phase will in general depend on it. 
In particular, as $m_z(0) \rightarrow 1$, one will have $p_+ = 1/2$, at $\gamma = \gamma_c(h)$, $p_+ = 1$ at $\gamma > \gamma_c (h)$. In the regime $\gamma \gg h$ instead, the Hamiltonian dynamics can be neglected on the whole phase space, leading to the single equation for $m_z$
\begin{equation} \label{eq:largegamma}
    \d m_z = \sqrt{\gamma} (1-m_z^2) \d \xi \ , 
\end{equation}
which preserves the ensemble average of $m_z$. We thus conclude that $\lim_{N \rightarrow \infty} \lim_{t \rightarrow \infty} \overline{m_{z}} = m_z (0)$.  As shown in App.~\ref{app:LargeGamma}, in this limit also the full counting statistics at any time $t$ can be computed exactly.

\section{Conclusion \& Outlook} \label{s:conclusions}

In this work, we explored the dynamics of the LMG model under continuous weak monitoring. We derived a set of noisy semi-classical SDEs (Eq.~\eqref{eq: SHE large S}) for the expectation values of the magnetization of the system, valid in the thermodynamic limit $N \rightarrow \infty$. Our findings highlight a critical distinction, as the large-$N$ limit does not commute with the long-time limit. While for any finite $N$ the system asymptotically approaches a trivial steady state under noise averaging, in the thermodynamic limit the system exhibits a non-trivial asymptotic behavior, which is already visible at the level of ensemble averages of the expectation values of the magnetization. 
We discovered that for small enough magnetic fields, such quantities exhibit a MIPT driven by the measurement strength. Our theoretical analysis was able to provide a full quantitative characterization of this transition, which can be explained in terms of a bifurcation of the stationary distribution in the thermodynamic limit, as the system develops two different attractive absorbing states. Our predictions are corroborated by numerical results.  

This study paves the way for further exploration of noise-induced phenomena in other quantum many-body systems and highlights the potential for using noise as a tool for probing and controlling quantum states. In particular, by demonstrating that continuous monitoring can induce non-trivial effect in fully-connected models already at the level of ensemble averages, our findings provide new insights into the interplay between monitoring and quantum dynamics in long-range interacting systems. Although for finite $N$ this behavior is only visible on an intermediate timescale $1 \ll t \ll t_{\rm Ehr}$, since $t_{\rm Ehr}$ grows algebraically in $N$, these timescale can become physically accessible. Moreover, as our order parameter $p_{+}$ is linear in the quantum state, the possibility of an experimental realization, e.g. in the context of cavity setups, is not hindered by postselection issues. 

Further investigations are needed in order to understand whether such behaviors are robust or in presence of short-range couplings, which could be obtained by means of a spin-waves analysis~\cite{lerose2018chaotic,delmonte2024monitored}. Analogously, it is worthwhile to investigate whether our theoretical picture still holds if we replace our all-to-all model with slowly-decaying power-law decaying couplings (i.e. strong-long range interactions~\cite{DEFENU20241}), as these are known to give rise to a phenomenology which can be similar to the fully-connected case up to thermodynamically large timescales~\cite{defenu2021metastability,pappalardi2018scrambling,giachetti2023entanglement,kastner2011diverging}. 

Finally, while our analysis focuses on the monitoring of a collective observable, which preserves the global symmetry of the LMG model, the impact of local measurements on the dynamics of open, fully-connected, quantum systems might reveal an interesting interplay between local and mean-field effects. \\

\begin{acknowledgments}
We acknowledge the use of QuTiP for the numerical simulations~\cite{qutip}. We are particularly grateful to Rosario Fazio, Andrea de Luca, and Marco Baldovin for inspiring discussions on the topic. This work was supported by the PNRR MUR project PE0000023-NQSTI, and the PRIN 2022 (2022R35ZBF) - PE2 - “ManyQLowD”.
\end{acknowledgments}

\bibliography{biblography}

\clearpage
\appendix

\section{Derivation of Eq.\,\eqref{eq: SSE}}\label{app:weak_monitoring}

In this appendix we introduce the concept of weak-measurements and derive the SSE for continuously monitored quantum dynamics ~\cite{Wiseman_Milburn_2009,Jacobs_2006}.

\subsection{Weak-measurements}

A weak measurement is a measurement that extracts partial information from a quantum system. 
The traditional way of describing a quantum measurement, referred to as von Neumann projective measurements, is to write the state of the system into the eigenstates of a given observable $O$, namely\begin{align}
    \ket{\psi} = \sum_a c_a \ket{a}\,\,,\,\,\,\,  O = \sum_n o_a \dyad{a}.
\end{align}
The probability $P(a)$ of measuring $o_a$ and thus projecting the state of the system into $\ket{a}$ is $|c_a|^2$. The state of the system state therefore transforms as
\begin{equation}
\varrho = \dyad{\psi} \longrightarrow \varrho_f = \dyad{a} = \frac{\Pi_a \varrho \Pi_a}{\Tr{\Pi_a \varrho \Pi_a}},
\end{equation}
with $P(a) = \Tr{\Pi_a \varrho \Pi_a} = |c_a|^2$. Such a measurement leaves completely projects $\varrho$ in an eigenstate of the observable, thus extracting maximal information.

Generalized measurements are described in terms of POVMs, i.e. positive operator-valued measures. Consider a set of operators $L_a$ such that $\sum_a L_a^\dagger L_a = \mathbb{I}$. The measurement process can be described in a similar fashion by transforming\begin{equation}
    \varrho \longrightarrow \varrho_f = \frac{L_a \varrho L_a^\dagger}{\Tr{L_a \varrho L_a^\dagger}}
\end{equation}
with probability $P(a) = \Tr{L_a \varrho L_a^\dagger}$. 
%
The following simple model provides a clear illustration of a weak projective measurement. Consider a two-level ancilla, represented by the eigenstates $\{\ket{+},\ket{-}\}$ of the Pauli matrix $\sigma_z$, initially prepared in the state\begin{equation}
    \ket{a} = \frac{\ket{+}+\ket{-}}{\sqrt{2}}\,.
\end{equation}
The ancilla is coupled to the system of interest, represented by the state $\ket{\psi_t}$. Let both the ancilla and the system evolve over a time $\Delta t$ under the unitary evolution operator,  $\hat{U}_{S+A}(\Delta t)$ \begin{equation}
    \hat{U}_{S+A}(\Delta t) \ket{\psi_{t}}\ket{\alpha} = \left(L_{+}\ket{\psi_t}\right)\ket{+}+\left(L_{-}\ket{\psi_t}\right)\ket{-}, 
\end{equation}
where $L_{\pm}= \mel{\pm}{\hat{U}_{S+A}(\Delta t)}{a}$ act exclusively on the system's Hilbert space. Following this evolution, a projective measurement acts on the ancilla along the $z$-axis, resulting in the outcome $a = \pm 1$. As a result, the back-action of the measurement places the system in the state
\begin{equation}
    \ket{\psi_{t+\Delta t}} = \frac{L_{a}\ket{\psi_t}}{\sqrt{\mel{\psi_t}{L_{a}^\dagger L_{a}}{\psi_t} }}\, .
\end{equation}

\subsection{Continuous limit}
Let us now consider the continous limit $\Delta t \to 0$. We have to derive the explicit form of the operators $L_{\pm}$. If we want to measure the observable $X$, let us consider the coupling between $\mathcal{S} + \mathcal{A}$ of the form
\begin{equation}
    H_{S + A} =  H + \lambda X \sigma_y \;. 
\end{equation}
We now take the limit $\Delta t \to 0$, scaling lambda in such a way that $\gamma = \lambda^2 \Delta t$ is kept constant.  Expanding the propagator $U$ we have
\begin{equation}
\begin{split}
    &U_{S+A} = e^{- i (\Delta t H + \sqrt{\gamma \Delta t} X \sigma_y)} \\ &= 1 - i \Delta t H - i \sqrt{\gamma \Delta t} X \sigma_y - \frac{1}{2} \gamma \Delta t O^2  + O (\Delta t^{3/2})
\end{split}
\end{equation}
We thus obtain for the $L_{\pm}$
\begin{equation}
    L_{\pm} = \frac{1}{\sqrt{2}} \left(1 - \imath \Delta t H \mp \sqrt{\gamma \Delta t} X - \frac1 2 \gamma \Delta t X^2  \right) \,.
\end{equation}
In order to compute the norm, we expand
\begin{equation}
    L_{a}^\dagger L_{a} = \frac 12 - a \sqrt{\gamma \Delta t} X \, , 
\end{equation}
while the corresponding probabilities become
\begin{equation}
    P(a) = \frac 12 - a \sqrt{\gamma \Delta t} \mean{X} \,.
\end{equation}
It follows that
\begin{equation}
    \begin{split}
    \ket{\psi_{t+ \Delta t}} &= \ket{\psi_t} - i H \Delta t \ket{\psi_t} - a \sqrt{\gamma \Delta t} ( X - \langle X \rangle) \ket{\psi_t} \\ &+ \frac 3 2  \gamma \Delta t \langle X \rangle^2 \ket{\psi_t}  - \gamma \Delta t \langle X  \rangle X \ket{\psi_t}- \frac 1 2 \gamma \Delta t X^2 \ket{\psi_t} 
    \end{split}
\end{equation}
Finally, we have to put this in the form of a stochastic equation. To this aim we observe that the measurement outcome $a$ is a random variable that satisfies
\begin{equation}
    \overline{a} = - 2 \sqrt{\gamma \Delta t} \langle X \rangle  \;, \quad \overline{a^2} = 1 \;.
\end{equation}
Let us thus introduce $Y_t =  \sqrt{\Delta t} \sum_{t'\leq t} a$: in the limit $\Delta t \to 0$ this converges to a continous stochastic variable such that
\begin{equation}
\label{eq:dY}
    \d Y = - 2 \sqrt{\gamma} \langle X \rangle \d t + \d \xi \,,
\end{equation}
where $\xi_t$ is a real Wiener process (i.e. $\overline{\d \xi} = 0$ and $\overline{\d \xi^2} = \d t$). 
Replacing $a \to \d Y/\sqrt{\Delta t}$ and using \eqref{eq:dY} in the limit $\Delta t \to 0$, we recover 
\begin{equation}
\begin{split}
    \d\ket{\psi} &= - i H \d t \ket{\psi_t} + \\ &\left(\sqrt{\gamma} ( X - \langle X \rangle ) \d \xi - \frac \gamma 2 ( X - \langle \hat X \rangle)^2 \d t \right) \ket{\psi_t}
\end{split}
\end{equation}

\section{Coherent spin states}
\label{app:coherentstates}
\noindent
The so-called spin, or $SU(2)$, coherent states are the generalization of the usual coherent states of the harmonic oscillator to angular momentum states. Like for the harmonic oscillator, they are a complete but not orthogonal set of states and they minimize Heisenberg’s uncertainty relations \cite{radcliffe1971, arecchi1972, perelmov1977}. 
Let us consider a single spin $s$ (in our case, $s=N/2$). For every spatial direction $\hat{\mathbf{n}}= \left( \sin \theta \cos \phi, \sin \theta \sin \phi, \cos \theta \right)$ there is an associated coherent spin state defined as
\begin{equation}
\ket{\Omega_{\theta,\phi}} := e^{-i \theta ( \bold{\hat{n}} \cross \hat{\bold{z}})\cdot \bold{S}} \ket{s} ~ ,
\end{equation}
which is also the maximum eigenstate of $\hat{\bold{S}} \cdot \hat{\bold{n}}$. Their expectation values of $\hat{S}_\alpha$ have a clear geometric interpretation, as they are equal to the projections 
\begin{equation}
   \langle S_\alpha \rangle \equiv \langle \Omega_{\theta \phi} | S_\alpha |\Omega_{\theta \phi} \rangle = n_\alpha S\,.
\end{equation}
Such result can be found by expanding in terms of the eigenbasis of $S_z$ $\ket{m}$ ($m= -s, \cdots, s$) as
\begin{align*}
\ket{\Omega_{\theta \phi}}= \sum_{m=-s}^{s} { \sqrt{{2s \choose m+s}} \left( e^{i\phi} \sin{\frac{\theta}{2}}\right)^{s-m} \left( \cos{\frac{\theta}{2}}\right)^{s+m} \ket{m} }.
\end{align*}  
An important property they satisfy is $\langle \hat{m}_\alpha \hat{m}_\beta\rangle = \langle \hat{m}_\alpha \rangle \langle \hat{m}_\beta\rangle + O(1/S)$. Below we derive it with the explicit higher-order correction that is necessary for describing quantum trajectories in the semiclassical limit.

\subsection{Derivation of Eq.\,\eqref{eq: mean_field_correction}}

We now need to compute $\mean{S_z^q} = \bra{\Omega_{\theta \phi}} S_z^q \ket{\Omega_{\theta \phi}}$. Expanding again over $|m\rangle$ one finds 
\begin{align}
\label{eq: moments}
  \mean{S_x^q} &=  \sum_{m=-s}^{s} m^q \  {2s \choose m+s} \times \\  &\times\left(  \frac{1-\cos \theta}{2} \right)^{s-m} \left(  \frac{1+\cos \theta}{2} \right)^{s+m} \equiv \mean{m^q}_b  , \notag
\end{align}
where $\mean{\cdot}_b$ represents the average over a binomial distribution, centered in zero with $p,q = (1 \pm \cos \theta)/2$. Thus we have
\begin{align}
      \mean{S_z} = &\mean{m}_b = s(p-q) = s \cos \theta \ , \\
    \mean{S^2_z} = &\mean{m^2}_b = \mean{m}_b^2 + 2 s pq = \mean{S_z}^2 + \frac{s}{2} \sin^2 \theta \, , \notag
\end{align}
and analogously $\mean{S_x} = s \sin \theta \cos \phi,\,\mean{S_y} = s \sin \theta \sin \phi$\, .

Expectation values like $\langle S_x S_z + S_z S_x\rangle$ can be calculated analogously if we employ
\begin{equation}
\langle m|S_x|n\rangle = \frac{1}{2}\sqrt{s(s+1)-mn}\,\bigl(\delta_{m,n+1}+\delta_{m,n-1}\bigr)\,,\\    
\end{equation} 
by expanding in the $S_z$ basis we get
\begin{widetext}
\begin{equation}
\begin{split}
    \langle &S_x S_z + S_z S_x\rangle = 
  \sum_{m,m'} \sqrt{{2s\choose{s+m}}{2s\choose{s+m'}}}\,e^{i(m'-m)\phi}\left(\sin\frac{\theta}{2}\right)^{2s-m-m'}\left(\cos\frac{\theta}{2}\right)^{2s+m+m'} \times \\
    & \times \frac{1}{2}(m+m')\sqrt{s(s+1)-mm'}\,\bigl(\delta_{m',m+1}+\delta_{m',m-1}\bigr) \\
    & = \sum_{m=-s}^s \sqrt{{2s\choose{s+m}}{2s\choose{s+m}}}\,(2m+1)\sqrt{s(s+1)-m(m+1)} \ \frac{e^{i\phi}+e^{-i\phi}}{2}  \left(\sin\frac{\theta}{2}\right)^{2(s-m)-1} \left(\cos\frac{\theta}{2}\right)^{2(s+m)+1} \\
    & = \cos\phi\cot{\frac{\theta}{2}} \sum_{m=-s}^s {2s\choose{s+m}}(s-m)(2m+1)\left(\frac{1-\cos \theta}{2}\right)^{s-m}\left(\frac{1+\cos \theta}{2}\right)^{s+m} \\
    &=  \cos\phi\cot{\frac{\theta}{2}} \bigl(s+(2s-1)\langle m \rangle_b -2\langle m^2\rangle_b \bigr)\\ 
    &= s(2s-1) \cos\theta\sin\theta\cos\phi + O(1)\, ,
\end{split}
\end{equation}
\end{widetext}
where we have used $\sqrt{s(s+1)-m(m+1)}=\sqrt{(s-m)(s+m+1)}$ and ${{2s}\choose{s+m+1}}={{2s}\choose{s+m}} \frac{s-m}{s+m+1}$. 
In terms of $\mean{S_\alpha}$ we thus get
\begin{equation}
\begin{split}
    \langle S_x S_z + S_z S_x\rangle &= (2-s^{-1}) \langle S_x\rangle \langle S_z\rangle + O(s^{-2})\,,\\
    \langle S^2_z \rangle &= \langle S_z \rangle^2 + \frac{s}{2}  \left(1  - \frac{\langle S_z \rangle^2}{s^2} \right) \, , 
\end{split}    
\end{equation}
or equivalently 
\begin{align}
    \langle \hat{m}_x , \hat{m}_z\rangle_c &=  - s^{-1} \langle \hat{m}_x \rangle \langle \hat{m}_z \rangle + O(s^{-2})\,,\notag \\
     \langle \hat{m}_z , \hat{m}_z\rangle_c &= s^{-1}  \left(1  - \mean{\hat{m}_z}^2 \right) \, .   
\end{align}
The result for the other pairs of components can be easily obtained by cyclic permutation of the indices, so that Eq.\,\eqref{eq: mean_field_correction} follows. 
\begin{figure}
    \centering
    \includegraphics[width=.75\linewidth]{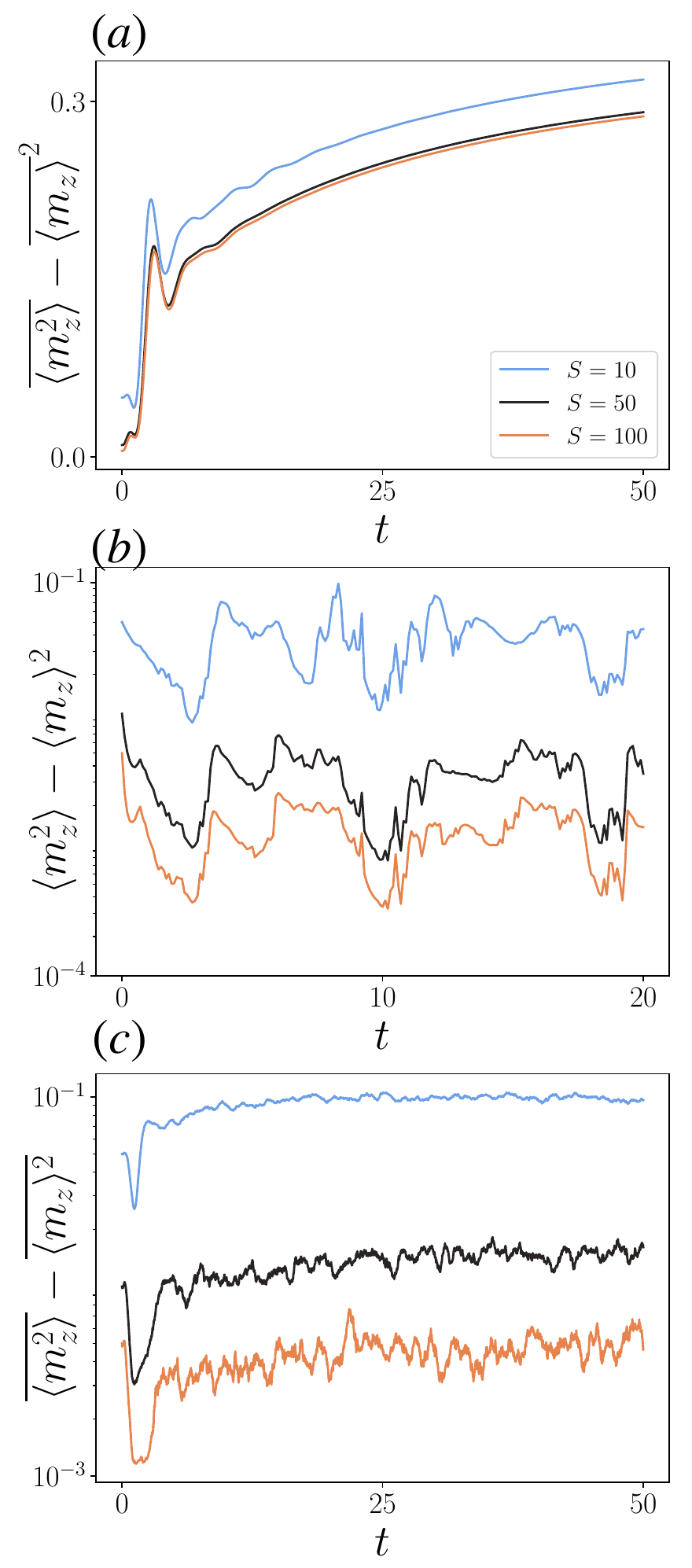}
    \caption{(a) Evolution of $\overline{\mean{m_z^2}}-\overline{\mean{m_z}}^2$ evaluated over the mean state; both $\overline{\mean{m_z^2}}$ and $\overline{\mean{m_z}}$ are linear in the average state. (b) Validation of the semiclassical approximation along individual quantum trajectories. We show that ${\mean{m_z^2}-\mean{m_z}^2}$ for increasing $S$ tends to zero uniformly in time, as in Eq.~\eqref{eq: large S approx}. The figure refers to the same trajectory, i.e. same realization of the stochastic process $\d \xi_t$, for different $S$. The initial condition is $m_y(0)=m_z(0)=0$, $m_x(0)=1$, for $h=0.5$, $\gamma=0.1$. (c) As in panel (b) averaged over the set of quantum trajectories. }
    \label{fig:confronto_lindblad_trajectories}
\end{figure}

\section{Semiclassic limit and ensemble averages}
\label{app: semiclassical approximation}

Here, we present numerical evidence to support the semiclassical approximation as described in Eq.~\eqref{eq: large S approx}. Specifically, in Fig.~\ref{fig:confronto_lindblad_trajectories} we show that
\begin{equation}
\label{eq: approx}
    {\mean{\hat{m}_z^2}} = {\mean{\hat{m}_z}^2} + O(1/S)
\end{equation}
along individual quantum trajectories. This approximation is not valid if the mean is computed on average state.
Averaging over quantum trajectories Eq.~\eqref{eq: approx} provides 
\begin{equation}
    \langle \hat{m}_z^2\rangle_\rho = \overline{\langle \hat{m}_z^2\rangle} = \overline{\langle{\hat{m}_z\rangle}^2} + O(1/S)\,,
\end{equation}
whereas the decoupling applied to $\rho$ would give
\begin{equation}
    \langle \hat{m}_z^2\rangle_\rho \overset{?}{=} {\langle \hat{m}_z\rangle_\rho}^2 + O(1/S) {=} \overline{\langle{\hat{m}_z\rangle}}^2 + O(1/S)\,.
\end{equation}
As illustrated in Fig.~\ref{fig:confronto_lindblad_trajectories}, $\langle \hat{m}_z^2\rangle_\rho - {\langle \hat{m}_z\rangle_\rho}^2$ does not converge to zero while it does for individual trajectories and their averages. This means
that classical correlations among quantum trajectories cannot be neglected as $\overline{m_z^2}\neq\overline{m_z}^2$.

\section{Hamiltonian dynamics close to the separatrix} \label{app:separatrix}
\noindent
We will now give an alternative explanation to our results for the LMG model. First, let us notice that the separatrix corresponds to $\mathcal{H} (m_z,\phi) = - 2h$. If we thus consider energies $ E =  - 2h + \epsilon$, with $\epsilon$ small, we get 
\begin{equation}
    m_z(\phi) = \frac{1}{ \cos^2 \phi} \left( h - \sqrt{(h- \cos^2 \phi)^2 +  \epsilon \cos^2 \phi} \right) 
\end{equation}
from which we see that the closest value to the barrier $m_z=1$ is in corespondence of $\phi = \pi/2$, namely 
\begin{equation}
    m_z( \pi/2) = 1 - \frac{\epsilon}{2h} \ . 
\end{equation}
Let us assume that at $t=0$ the system in in $m_z(\pi/2)$ so that $\Delta z(t=0) \propto \epsilon$. For amll $\Delta z$ the typical timescale $T(\Delta z)$ spent by a trajectory close to the $m_z=1$ absorbing wall, can be estimated as the half-period period of the trajectory with energy $E = -2h + \epsilon$, i.e. 
\begin{align}
 \notag   T(\Delta z) &= \frac{1}{2} \int^{\pi}_0 \frac{\d \phi}{\dot{\phi}} = \frac{1}{4} \int^{\pi}_0 \frac{\d \phi}{ m_z(\phi) \cos^2 \phi - h} \\&= \frac{1}{4} \int^{\pi}_0 \frac{\d \phi}{\sqrt{(h-\cos^2 \phi)^2 + \epsilon \cos^2 \phi}}    \notag
\end{align}
which, through the substitution $u = \cos^2 \phi$ can be recast as
\begin{equation}
     T(\Delta z) = \frac{1}{8} \int_0^1 \frac{\d u}{\sqrt{u(1-u)}} \frac{1}{\sqrt{(u- h)^2 +\epsilon u}} \  . 
\end{equation}
For small $\epsilon$, only the region close to $u = h$ is going to contribute. By setting $u = h + \sqrt{\epsilon} \tilde{u}$ we have
\begin{align}
     T(\Delta z) &\approx \frac{1}{8 \sqrt{h(1-h)}} \int^{1/\sqrt{\epsilon}}_{-1/\sqrt{\epsilon}} \frac{\d \tilde{u}}{\sqrt{\tilde{u}^2 + h}} \\ \notag
     &\approx \frac{1}{4 \sqrt{h(1-h)}} (-\ln \epsilon) \approx \frac{1}{4 \sqrt{h(1-h)}} (-\ln \Delta z) \ . 
\end{align}
By introducing once again $\Delta z = e^{-s}$, one has: 
\begin{equation}
    T(e^{-s}) \sim  \frac{s}{4 \sqrt{h(1-h)}}
\end{equation}
As the Hamiltonian dynamics in $m_z$ critically slows down close to $m_z=1$, for $t < T(e^{-s})$. This happens for the trajectories such that 
\begin{equation} \label{eq:limitvelocity}
    s < 4 \sqrt{h(1-h)} t \ .
\end{equation}
As seen in the main text, however, the noisy dynamics close to the wall is given by
\begin{equation}
    \d s = 2 \gamma \d t + 2 \sqrt{\gamma} \d \xi
\end{equation}
i.e. a Gaussian distributed around $2 \gamma t$ with a typical width $\sim \sqrt{t}$. The probability of having such a trajectory that fulfills the conditions~\eqref{eq:limitvelocity} for large times is thus null if $ 2 \gamma > 4 \sqrt{h(1-h)} = 2 \gamma_c (h)$, so that in this range of values the barrier is repulsive. 

\section{Large $\gamma$ limit} 
\label{app:LargeGamma}
\noindent 
We will now address the limit of large $\gamma$. We start from Eq.~\eqref{eq:largegamma} 
\begin{equation}
    \d m_z = \sqrt{\gamma} (1-m_z^2) \d \xi
\end{equation}
It is convenient to set $m_z=\tanh(s)$, $\tau = \gamma t$, so that 
\begin{equation}
    \d s = \tanh(s) \d \tau + \d\xi 
\end{equation}
(with $\d \xi^2 = d \tau$). This can be rewritten as Fokker-Plank equation for the probability distribution $P(s,t)$ of the form 
\begin{equation}
    \partial_\tau P(s,\tau) = \partial_s \left( V^{\prime} (s) \ P(s,\tau) + \frac{1}{2} \partial_s P(s,\tau) \right)
\end{equation}
with $V= - \ln \cosh(s)$, $P(s,0) = \delta(s-s_0)$ and $s_0 = \arctanh z(0)$. By means of the substitution: 
\begin{align}
    \notag P(s,\tau) &= e^{-\tau/2} e^{-(V(s)- V(s_0))} \psi(s, \tau) \\&=  e^{-\tau/2} \frac{\cosh(s)}{\cosh(s_0)} \psi(s,\tau) \, 
\end{align}
with $\psi(s,0) = \delta(s-s_0)$, one has 
\begin{equation}
    \partial_{\tau} \psi (s, \tau) = \frac{1}{2} \partial_s^2 \psi (s, \tau)
\end{equation}
from which finally: 
\begin{equation}
    \psi (s, \tau) = \frac{1}{\sqrt{2 \pi \tau}} e^{-(s-s_0)^2/2 \tau} \, . 
\end{equation}
Coming back to $P(s,\tau)$ we have
\begin{align}
    P(s, \tau) = \frac{1}{2 \cosh s_0} \frac{1}{\sqrt{2 \pi \tau}} \bigl( e^{-s_0} &e^{-(s-s_0 + \tau)^2/2 \tau} +\\+ e^{s_0} &e^{-(s-s_0 - \tau)^2/2 \tau} \bigr) \ ,  
\end{align}
which is the superposition of two Gaussian distributions traveling to the left and the right respectively, with weights that depends on the initial condition. 

From this, one can recover $p_{\pm}$ and the ensable average of $m_z$. For example
\begin{equation}
    p_{+} = \lim_{\tau \rightarrow \infty} \int^{\infty}_0 \d s \ P(s,\tau) = \frac{e^{s_0}}{e^{s_0} + e^{-s_0}} = \frac{1 + z(0)}{2}
\end{equation}
and analogously for $p_-$, so that we recover
\begin{equation}
    \lim_{N \rightarrow \infty} \lim_{t \rightarrow \infty} \overline{m_{\alpha}} = p_+ - p_- = z(0) 
\end{equation}

\end{document}